\documentclass[12pt]{article}

\usepackage{mathtools}
\usepackage{booktabs}
\usepackage[english]{babel} 
\usepackage[protrusion=true,expansion=true]{microtype} 
\usepackage{amsfonts,amsthm}
\usepackage{amsmath}
\usepackage[ruled,vlined,algo2e]{algorithm2e}
\usepackage{ amssymb }
\usepackage{graphicx}
\usepackage{array}
\usepackage{bbm}
\usepackage{multirow}
\usepackage{hhline}
\usepackage[titletoc]{appendix}
\usepackage[dvipsnames]{xcolor}
\usepackage{xr}
\usepackage[symbol]{footmisc}

\usepackage{booktabs} 
\usepackage{setspace}
\usepackage{microtype} 
\usepackage{subcaption}
\usepackage{tabularx}
\usepackage[font = small,labelfont=bf,textfont=it]{caption} 
\usepackage{algorithm}
\usepackage[noend]{algpseudocode}

\makeatletter
\newcommand{\bm}[1]{\mathbf{#1}}
\newcommand{\bi}{\begin{itemize}}
\newcommand{\ei}{\end{itemize}}
\newcommand{\distas}[1]{\mathbin{\overset{#1}{\kern\z@\sim}}}

\newsavebox{\mybox}\newsavebox{\mysim}

\theoremstyle{definition}

\newcommand{\distras}[1]{%
  \savebox{\mybox}{\hbox{\kern3pt$\scriptstyle#1$\kern3pt}}%
  \savebox{\mysim}{\hbox{$\sim$}}%
  \mathbin{\overset{#1}{\kern\z@\resizebox{\wd\mybox}{\ht\mysim}{$\sim$}}}%
}
\newcolumntype{C}[1]{>{\centering\let\newline\\\arraybackslash\hspace{0pt}}m{#1}}

\DeclareMathOperator*{\argmax}{\arg\!\max}

\newcommand{\blind}{1}

\makeatletter
\newcommand\footnoteref[1]{\protected@xdef\@thefnmark{\ref{#1}}\@footnotemark}
\makeatother

\usepackage{ragged2e}
\def\spacingset#1{\renewcommand{\baselinestretch}%
{#1}\small\normalsize} \spacingset{1}

\usepackage{natbib}

\begin{document}


\if1\blind
{
\title{TSEC: a framework for online experimentation under experimental constraints}
\author{Simon Mak\footnote{Joint first authors}\;\footnote{Department of Statistical Science, Duke University}\;, \; Yuanshuo Zhao$^*$\footnote{Uber Technologies Inc.},\; Lavonne Hoang$^{\text{\textdagger}}$, \; C. F. Jeff Wu\footnote{H. Milton Stewart School of Industrial \& Systems Engineering, Georgia Institute of Technology}\;\footnote{This research is supported by NSF DMS-1914632 and NSF CSSI Frameworks 2004571.}}


 \maketitle
} \fi

\if0\blind
{
  \bigskip
  \bigskip
  \bigskip
  \begin{center}
    {\Large\bf TSEC: a framework for online experimentation under experimental constraints}
\end{center}
  \medskip
} \fi

\bigskip

\begin{abstract}
Thompson sampling is a popular algorithm for solving multi-armed bandit problems, and has been applied in a wide range of applications, from website design to portfolio optimization. In such applications, however, the number of choices (or arms) $N$ can be large, and the data needed to make adaptive decisions require expensive experimentation. One is then faced with the constraint of experimenting on only a small subset of $K \ll N$ arms within each time period, which poses a problem for traditional Thompson sampling. We propose a new Thompson Sampling under Experimental Constraints (TSEC) method, which addresses this so-called ``arm budget constraint''. TSEC makes use of a Bayesian interaction model with effect hierarchy priors, to model correlations between rewards on different arms. This fitted model is then integrated within Thompson sampling, to jointly identify a good subset of arms for experimentation and to allocate resources over these arms. We demonstrate the effectiveness of TSEC in two problems with arm budget constraints. The first is a simulated website optimization study, where TSEC shows noticeable improvements over industry benchmarks. The second is a portfolio optimization application on industry-based exchange-traded funds, where TSEC provides more consistent and greater wealth accumulation over standard investment strategies.


\end{abstract}

\noindent%
\vfill

\newpage

\spacingset{1.5}

\section{Introduction}
The multi-armed bandit (MABs; \citealp{2015arXiv151000757B}) describes a sequential experimental procedure where, at each time, one chooses to perform an experiment (or several experiments) from a set of $N$ possible choices. Each experiment generates a random reward, and the goal is to adaptively choose a sequence of experiments which maximizes the accumulated reward. The term ``multi-armed bandit'' originates from gambling, where a player can play different arms on a row of slot machines; each of the $N$ experimental choices is then referred to as an ``arm''. MABs have been broadly used in a variety of applications, including clinical trials \citep{gittins2011multi}, website optimization \citep{white2012bandit}, and financial portfolio design \citep{shen2015portfolio}.

Given the wide applicability of this framework, MABs have received significant attention in recent years, particularly in the statistics and machine learning community. Existing MAB methodology can be roughly split into two categories: methods for \textit{regret minimization}, which aim to maximize an experimenter's accumulated reward (or equivalently, minimize an experimenter's accumulated regret), and methods for \textit{best-arm identification}, which aim to identify the arm yielding greatest reward. For regret minimization, early seminal papers include \cite{Lai:1985:AEA:2609660.2609757} and \cite{Katehakis8584}, who developed optimal finite-time sampling strategies for regret minimization. Such strategies are, however, computationally intractable, and recent work has focused on efficient approximations. This includes the $\epsilon$-greedy sampling methods \citep{Sutton1998ReinforcementLA}, the upper confidence bound (UCB) strategies \citep{pmlr-v23-agrawal12}, and the widely-used \textit{Thompson sampling} (TS) method \citep{Russo:2018:TTS:3283245.3283246}. The latter method was originally proposed in \cite{10.1093/biomet/25.3-4.285}, then popularized by two recent papers \citep{Scott:2010:MBL:1944422.1944432,chapelle2011empirical}. For best-arm identification, \cite{domingo2002adaptive} and \cite{even2006action} explored the problem of minimal sampling times for identifying an $\epsilon$-optimal arm, and \cite{audibert2010best} proposed a UCB-based policy which is nearly asymptotically optimal. Building on these works, there is now a rich body of literature on both topics.

One limitation of standard MAB methods, however, is that it presumes an experimenter can play \textit{any} of the $N$ arms within each time period. This may not be possible in problems where experimentation is costly or time-intensive, particularly when the number of considered arms $N$ is large as well. Consider, for example, the problem of website optimization \citep{Ash:2012:LPO:2361815}, where the goal is to design a website which maximizes the conversion percentage of website visitors to customers. Here, arms represent websites with different combinations of factors, say, font family, font size, image location, and background color. If each of these four factors has five choices (or levels), then there are $N=5^4=625$ possible experimental arms. In practice, however, a company typically cannot test more than $K=30$ website versions simultaneously due to maintenance costs. This results in a so-called ``arm budget constraint'', where the number of available arms at a given time ($K$) is less than the number of considered arms ($N$). Similarly, for portfolio optimization \citep{black1992global}, an investor can only invest in a limited number of portfolios due to transaction costs. Given this constraint, an experimenter would need to jointly (i) identify a good subset of $K$ arms for experimentation, and (ii) allocate experimental resources over these arms to maximize reward. There is much literature on addressing (ii), but not the joint problem of (i) and (ii) -- we aim to tackle this joint problem in this work.



We propose a new method called \textbf{T}hompson \textbf{S}ampling under \textbf{E}xperimental \textbf{C}onstraints (TSEC), which addresses this arm budget constraint for MABs. TSEC makes use of a Bayesian interaction model\footnote{For simplicity, we focus on Bernoulli bandits in this work, where rewards are binary variables indicating successes or failures. The proposed TSEC methodology can be generalized in a straightforward manner for a broader class of reward distributions.} to model correlations between reward distributions over different arms, with prior distributions on effect coefficients to enforce the principle of effect hierarchy (see \citealp{Joseph2006ABA,WuHamada2009}). This fitted Bayesian model can then be used to jointly tackle the aforementioned goals (i) and (ii). For (i), the proposed model is integrated within a UCB best-arm identification framework, which allows an experimenter to sequentially add good arms and remove bad arms from the active set of arms, a procedure we call ``switching''. For (ii), the posterior samples from the model can be used within a modified Thompson sampling procedure to allocate experimental resources over selected arms. The key novelty in TSEC is the use of a Bayesian MAB framework to model correlations between arms, and the leveraging of this correlation for adaptive arm selection and experimentation. This allows TSEC to identify potentially good arms for experimentation, even if no data has been observed on such arms. In contrast, standard MAB methods typically assume that rewards are independently distributed between different arms \citep{gupta2019multi}. When the number of arms is large (i.e., $N \gg 1$), such methods require experiments over all arms for best-arm identification, which can be prohibitively costly. We will apply the TSEC methodology to the two motivating problems for website optimization and portfolio optimization, and show that it provides noticeable improvements over standard industry methods.



This paper is structured as follows. Section 2 describes the problem formulation and reviews the MAB, Thompson sampling, and arm budget constraints. Section 3 describes the proposed TSEC modeling framework and algorithm. Section 4 presents a simulated website optimization application and compares TSEC to industry standards. Section 5 applies TSEC to a portfolio optimization problem on industry-based exchange-traded funds (ETFs). Section 6 concludes the work.


\section{Problem formulation}

\subsection{Multi-armed bandits}
%
%
%
%


We first review the employed MAB framework, following \cite{bubeck2012regret} with slight modifications. Let $\mathcal{A} = \{1, \cdots, N\}$ denote the set of arms which can be experimented on, with $N$ being the total number of arms. For an arm $a \in \mathcal{A}$, a binary reward $Y_a \in \{0,1\}$ can be obtained via experimentation on arm $a$. It is typically assumed that:
\begin{equation}
Y_a | \mu_a \distas{indep.} \text{Bernoulli}(\mu_{a}), \quad a \in \mathcal{A}.
\label{eq:bern}
\end{equation}
Here, $\mu_{a}$ is the probability of a success (i.e., the expected reward). Suppose the experiment is conducted over a time frame of $T$ time periods. Within time period $t$, an experimenter then performs $n_{a,t}$ experiments at arm $a$, for a total of $n = \sum_{a \in \mathcal{A}} n_{a,t}$ experiments in a given time period. For the motivating website optimization problem, $T$ may be the number of weeks for the conversion experiment, $n_{a,t}$ is the amount of website traffic (i.e., users) assigned to website $a$ in week $t$, and $n$ is the total traffic for each week.


Clearly, if the success probabilities $\mu_{a}$ are known for all arms $a \in \mathcal{A}$, then the optimal strategy is to always pick the arm with largest probability. However, these probabilities are typically not known in practice. It would therefore be useful to define a measure of {regret} for making suboptimal decisions. The \textit{regret} $L_t$ at time $t$ is defined as:
\begin{equation}
L_t = \sum_{a \in \mathcal{A}} n_{a,t}(\mu^* - \mu_a).
\end{equation}
where $\mu^* = \max_{a \in \mathcal{A}} \mu_a$ is the expected reward of the optimal arm. In words, $L_t$ quantifies the expected loss of reward resulting from a suboptimal selection of arms. The expected \textit{reward} at time $t$ can similarly be defined as $R_t = \sum_{a \in \mathcal{A}} n_{a,t}\mu_a$. From this, one can then define the \textit{cumulative regret} $L$ over the full $T$ time periods as:
\begin{equation}
L = \sum_{t = 1}^T L_t,
\label{eq:basicregret}
\end{equation}
with $R = \sum_{t = 1}^T R_t$ its corresponding cumulative reward. In this paper, we focus on the objective of finding an allocation $(n_{a,t})_{a,t}$ which minimizes the cumulative regret $L$ (or equivalently, maximizing cumulative reward $R$), given the arm budget constraints described at the end of the section.

\subsection{Thompson sampling}
\label{sec:ts}

Thompson sampling is an increasingly popular sequential sampling strategy for regret minimization. It was first proposed in \cite{10.1093/biomet/25.3-4.285}, then popularized much later by two papers \citep{Scott:2010:MBL:1944422.1944432,chapelle2011empirical} which demonstrated its empirical performance. This sampling procedure is quite simple and intuitive from a Bayesian perspective, but powerful. The key steps are as follows. First, a modeler assigns independent priors for the reward probabilities on each arm. Next, the posterior distribution is updated using the reward data observed on each arm. Finally, one samples from the posterior of these probabilities to allocate experimental resources amongst the $N$ arms. The last two steps are then repeated for each time step until the experiment ends.

Algorithm \ref{alg:TPS} outlines the Thompson sampling algorithm under independent (uniform) $\text{Beta}(1,1)$ priors on the reward probabilities $(\mu_a)_{a \in \mathcal{A}}$ for each arm $a \in \mathcal{A}$, assuming $n$ experiments are performed within each time period. At each time period $t$, the arm played for experiment $j$ is chosen by first sampling one draw from the posterior distribution $[\mu_a|\text{data}]$ for each arm $a$, then selecting the arm $a^*$ with greatest sampled probability for experimentation. The posterior distribution for $\mu_{a^*}$ is then updated using the observed binary reward. This amounts to simply incrementing the Beta parameter $\alpha_{a^*}$ by 1 if $Y_{a^*}$ is a success and $\beta_{a^*}$ by 1 if it is a failure, since the Beta prior on $\mu_a$ is conjugate for the Bernoulli reward random variable. In this sense, the adaptive sampling procedure is driven by posterior sampling on the reward probabilities for each arm.


\begin{algorithm}[!t]
\caption{Thompson Sampling}\label{alg:TPS}
\vspace{-0.2cm}
\bi
\setlength{\itemsep}{-5pt}
\item For each arm $i = 1,...,N$, set the Beta prior parameters $\alpha_i = 1$ and $\beta_i = 1$.
\ei
\vspace{-0.2cm}
\For{\textup{each time period} $t = 1,2,...,T$}{%
\For{\textup{each experiment} $j = 1,2,...,n$}{%
\bi
\setlength{\itemsep}{-5pt}
\vspace{-0.2cm}
\item Sample $\theta_a$ from the posterior $\text{Beta}(\alpha_a,\beta_a)$, for each arm $a \in \mathcal{A}$.
\item Experiment on the arm $a^* = \argmax_{a \in \mathcal{A}} \theta_a$ with largest sampled posterior probability, and observe binary reward $Y_{a^*} \in \{0,1\}$.
\item Posterior update: If $Y_{a^*} = 1$, then set $\alpha_{a^*} = \alpha_{a^*} + 1$, else set $\beta_{a^*} = \beta_{a^*} + 1$.
\vspace{-0.2cm}
\ei
}
}
\end{algorithm}

Thompson sampling, despite being a simple procedure, enjoys many desirable theoretical properties, such as asymptotic optimality for regret minimization (see \citealp{pmlr-v23-agrawal12} and references thereafter). It is also intuitively appealing in that it inherently captures the well-known \textit{exploration-exploitation} trade-off in reinforcement learning \citep{kearns2002near}: exploration of the arm space $\mathcal{A}$, and exploitation of arms with high empirical reward probabilities. Note that, at the beginning of the sampling procedure, Thompson sampling encourages the \textit{exploration} of different arms in $\mathcal{A}$, since with little data, the posterior distributions of $(\mu_a)_{a \in \mathcal{A}}$ do not differ much between different arms. However, as more samples are collected, the posterior distribution of $(\mu_a)_{a \in \mathcal{A}}$ contracts towards the true reward probabilities via closed-form posterior updates, which then encourages \textit{exploitation} of arms with high empirical reward rates. Sub-optimal arms will then receive fewer and fewer experiments, which is as desired.\par


\subsection{Arm budget constraints}
\label{sec:constr}

As mentioned earlier, a key limitation of standard MAB methods (including Thompson sampling) is that it presumes an experimenter can play \textit{any} of the $N$ arms within a time period. When experiments are expensive and the number of considered arms $N$ is large, an experimenter may be restricted to testing only a subset of $K \ll N$ arms due to budget constraints. We call this the \textit{arm budget constraint}. For the motivating website optimization problem, this arises from several practical constraints: (i) website design costs may be expensive, and (ii) a minimum amount of traffic may be desired in each tested design to ensure reliable results and identify potential bugs (root causes for unusual traffic behavior).

Given this arm constraint, we will allow a total of $S$ ``switches" throughout the experimental procedure. During these switches, an experimenter has the opportunity to remove current arms which are likely suboptimal, and add new arms which may be promising. We will assume there are $T$ time periods between switches, which corresponds to a total experimental time frame of $ST$ time periods. To illustrate this set-up, consider the motivating website optimization problem. Here, a company can typically run batch conversion experiments in $T=8$ hourly time periods within a day (e.g., one batch per hour in a 9am - 5pm day). At the end of each day, the company can switch the current set of websites (i.e., arms), when traffic is at its lowest. These switches allow one to remove current websites which have low conversion rates, and add in new website designs based on previous experiments. With an experimental time frame of 14 days, this corresponds to a total of $S=14$ switches and $ST=14 \times 8$ time periods. 

We require a slight adaptation of regret for this new experimental framework with arm constraints. Within switch period $s$, let $\mathcal{A}_s \subset \mathcal{A}$ denote the ``active arm set'' -- the subset of arms which are selected for experimentation. Note that $\# \mathcal{A}_s = K$, which is typically much smaller than $\# \mathcal{A} = N$. Furthermore, let $n_{a,t,s}$ denote the number of experiments performed at arm $a$ in time period $t$ and switch period $s$. For arms $a$ which are not selected in switch $s$ (i.e.,  $a \notin \mathcal{A}_s$), $n_{a,t,s}$ must equal 0, since no experiments were performed. We then define the \textit{cumulative regret under arm constraints} as:
\begin{equation}
L = \sum_{s = 1}^S \sum_{t = 1}^T \sum_{a \in \mathcal{A}_s} n_{a,t,s} (\mu^* - \mu_a),
\label{eq:cumreg}
\end{equation}
where, as before, $\mu^* = \max_{a \in \mathcal{A}}\mu_a$ is the expected reward of the optimal arm. Compared to the earlier regret \eqref{eq:basicregret}, this new measure of regret accounts for the presence of the active arm set $\mathcal{A}_s$ within each switching period $s$. From \eqref{eq:cumreg}, we see that the minimization of regret under arms constraints requires two steps: (i) the selection of the active arm set $\mathcal{A}_s$, and (ii) the allocation of $n$ experimental runs amongst the arms in $\mathcal{A}_s$. We outline next the TSEC method, which aims to do both (i) and (ii) in a sequential manner.

\section{The TSEC methodology}
We now present the proposed Thompson Sampling under Experimental Constraints (TSEC) methodology. We first describe a Bayesian probit model with effect hierarchy priors for modeling correlations between Bernoulli rewards on different arms. The fitted correlations from this model allow TSEC to infer information on the many unobserved arms in the high-dimensional arm space $\mathcal{A}$, by leveraging observed data from the few sampled arms. We then describe a procedure which integrates this information for the joint goals of active arm selection and run allocation.

\subsection{Bayesian interaction model}


As noted earlier, a key limitation of standard MAB methods is that they assume rewards from different arms are independently distributed, and do not model for correlations between arms. In doing so, such methods do not yield meaningful inference on an arm unless experiments have been performed on this arm. This poses a problem in the current setting, where experiments are expensive and the number of arms is large. The following Bayesian interaction model addresses this.



Suppose the arm space $\mathcal{A}$ corresponds to an experiment with $M$ factors, where factor $m$ has $L_m$ levels, $m = 1, \cdots, M$. Each arm $a \in \mathcal{A}$ can then be represented by a vector $\bm{x}_a=(x_{a,1}, \cdots, x_{a,M})$, where $x_{a,m}$ denotes the level of the $m$-th factor. For example, in the earlier website optimization example with $M=4$ factors (font family, font size, image location, and background color), each with $L=5$ levels, the arm $a$ with level 1 for all factors can be represented as $\bm{x}_a=(1,1,1,1)$.

For an arm $a \in \mathcal{A}$ in this setting, we model its reward probability $\mu_a$ by the following probit interaction model:
\begin{equation}
\mu_a = \Phi\left( \beta_0 + \sum_{m=1}^M \beta_{x_{a,m}}^m + \sum_{m=1}^M \sum_{m'=1,m'<m}^M \beta_{x_{a,m},x_{a,m'}}^{m,m'} \right),
\label{eq:model}
\end{equation}
where $\Phi$ is the standard normal cumulative distribution function. Here, $\beta_0$ is an intercept term, $(\beta_l^m)_{l=1}^{L_m}$ are the main effects (MEs) for factor $m$, and ${(\beta_{l,l'}^{m,m'})_{l=1}^{L_m}}_{l'=1}^{L_{m'}}$ are the two-factor interaction (2FI) effects between factors $m$ and $m'$. Despite the slightly cumbersome notation, this is simply a two-way interaction model on the binary rewards over the arm space, using a probit link function. For identifiability, we further impose the baseline constraints (\citealp{WuHamada2009}, p.95) on MEs and 2FIs:
\begin{equation}
\beta_1^m = 0, \quad \beta_{l,1}^{m,m'} = \beta_{1,l'}^{m,m'} =0 , \quad \text{for all $m,m' = 1, \cdots, M$.}
\end{equation}
Of course, this modeling framework can be generalized in several ways: one can instead use a logistic or complementary log-log link function \citep{mccullagh2018generalized}, and can entertain higher-order interaction effects as well. Such modeling decisions can be made depending on the problem at hand.



Let $\boldsymbol{\beta}$ denote the vector of main effects and 2FI parameters in model \eqref{eq:model}. Note that, conditional on parameters $\boldsymbol{\beta}$, the reward random variables $Y_a$ and $Y_{a'}$ from two different arms $a$ and $a'$ are still independent by \eqref{eq:bern}. To model correlations between arms, we adopt a Bayesian approach and assign a prior model on $\boldsymbol{\beta}$. Since the number of arms $N$ (e.g., the number of website designs) can be large, this prior should ideally provide some structured regularization for the high-dimensional effect parameters. To this end, we adopt the following independent priors on $\boldsymbol{\beta}$:
\begin{equation}
\beta_0 \sim \mathcal{N}(0,\tau^2), \quad \beta_l^m \distas{i.i.d.} \mathcal{N}(0,\tau^2 r), \quad \beta_{l,l'}^{m,m'} \distas{i.i.d.} \mathcal{N}(0,\tau^2 r^2),
\label{eq:prior}
\end{equation}
where $r \in (0,1)$ and $\tau^2 > 0$ are hyperparameters. This can be seen as an extension of the functional prior in \cite{Joseph2006ABA} for multi-level factors. The intuition behind \eqref{eq:prior} is straight-forward and stems from the principle of \textit{effect hierarchy} \citep{WuHamada2009}: higher-order effects are assigned smaller prior variances, reflecting the belief that such effects are likely less important than lower-order effects. Such a prior would then allow for better inference of interaction terms between arms, particularly with limited experimental data.


The proposed model has some similarities, but also key differences with the fractional factorial (FF) bandits in \cite{Scott:2010:MBL:1944422.1944432}. Similar to our approach, the FF bandits makes use of a regression model for modeling the reward probabilities $\mu_a$. Their model, however, is much simpler in that it only considers one-way main effects. While this model has fewer parameters to estimate, it can yield poor optimization and regret minimization performance when the underlying relationship between arms has significant interaction terms (see, e.g., \citealp{Wea1990,2017ATM}). The proposed model accounts for these interaction terms, and regularizes such effects via the Bayesian effect hierarchy prior \eqref{eq:prior} to reduce the number of parameters to be estimated. Our approach also integrates this model for the target problem of arm budget constraints, which is not considered in \cite{Scott:2010:MBL:1944422.1944432}.

With this model in hand, the sampling of the posterior distribution $[\boldsymbol{\beta}|\text{data}]$ can be performed in a straight-forward manner. While the two-way layout probit model does not admit a closed-form Gibbs sampler, we can use a well-known data augmentation algorithm \citep{10.2307/2290350} for Markov Chain Monte Carlo (MCMC) sampling. This is implemented in the \textsf{R} package \texttt{MCMCpack} \citep{martin2011mcmcpack}. We further adopt a fully Bayesian implementation and sample the hyperparameters $\tau^2$ and $r$ under the weakly-informative hyperpriors $[\tau^2] \propto 1/\tau^2$ and $[r] \sim \text{Unif}[0,1]$. This yields a richer quantification of model uncertainty, which may translate to better active arm selection and regret minimization, as we shall see next. 

\subsection{Selection of active arm set}

Given posterior samples on $[\boldsymbol{\beta}|\text{data}]$, we now show how this can be used to adaptively select the active arm set $\mathcal{A}_s$ for experimentation. Conceptually, this is related to the aforementioned problem of best-arm identification (more specifically, the ``best-$K$ identification'' problem, see \citealp{jiang2017practical}) in MABs, where one aims to find the best arm (or the best $K$ arms) adaptively. We present below an extension of the popular UCB  best-arm identification framework \citep{audibert2010best} for active arm selection.

Consider first the posterior random variable $[\mu_a(\boldsymbol{\beta})|\text{data}]$, which quantifies posterior uncertainty on the reward probability of arm $a \in \mathcal{A}$, given observed data. A concentration of $[\mu_a(\boldsymbol{\beta})|\text{data}]$ on large values suggests arm $a$ has higher reward rates, whereas a concentration on small values suggests lower reward rates. Suppose the posterior distribution $[\mu_a(\boldsymbol{\beta})|\text{data}]$ is known for all arms $a \in \mathcal{A}$. Then one strategy for selecting the top $K$ arms would be to choose the arms $a$ with largest posterior $(1-\alpha)$-th quantile, i.e.:
\begin{equation}
\label{eq:ucbth}
\text{$K$-argmax}_{a \in \mathcal{A}} \;  Q_{1-\alpha}( [\mu_a(\boldsymbol{\beta})|\text{data} ]),
\end{equation}
where $\text{$K$-argmax}$ returns the largest $K$ arms, and $Q_{1-\alpha}(\cdot)$ returns the $(1-\alpha)$-th quantile. In words, we wish to choose the arms which yield the largest posterior $(1-\alpha)$-th upper confidence bound from the fitted Bayesian interaction model. Of course, the exact form of the posterior distribution $[\mu_a(\boldsymbol{\beta})|\text{data}]$ is unknown, so we can instead estimate the arm selection problem in \eqref{eq:ucbth} with its sample analogue:
\begin{equation}
\label{eq:ucb}
\mathcal{A}_s = \text{$K$-argmax}_{a \in \mathcal{A}} \; Q_{1-\alpha}( \{\mu_a(\boldsymbol{\beta}_p)\}_{p=1}^P),
\end{equation}
where $\boldsymbol{\beta}_1, \boldsymbol{\beta}_2, \cdots, \boldsymbol{\beta}_P$ are MCMC samples from the posterior distribution $[\boldsymbol{\beta}|\text{data}]$.

We note a key difference between the above arm selection approach and the standard UCB best-arm identification in \cite{audibert2010best}. One limitation for the latter is that it requires experimental runs on \textit{all} arms before a ``best'' arm can be conclusively selected, since the underlying probabilistic model does not account for correlations between different arms. For the current setting of $N \gg 1$ arms and expensive experimentation, this standard procedure can result in wasted experiments and slow identification of good arms. In contrast, our approach only requires experiments on a small set of arms, since it can then leverage the modeled correlations from the fitted Bayesian model to infer on arms which have yet to be experimented on. This allows for us to identify good arms for experimentation early on in the procedure.


\subsection{Allocation of experimental resources}

Having selected the set of active arms $\mathcal{A}_s$ for experimentation, consider next the problem of allocating the $n$ runs on these arms. Recall the Thompson sampling procedure in Section \ref{sec:ts}, where samples from the \textit{independent} posterior distributions of each arm are used to allocate runs. We can extend this idea for the proposed Bayesian framework by instead sampling from the posterior distribution $[\mu_a(\boldsymbol{\beta})|\text{data}]$. In particular, experimental runs are allocated by first sampling one draw $\boldsymbol{\beta}'$ from the posterior $[\boldsymbol{\beta}|\text{data}]$, then selecting the arm $a$ in the active set $\mathcal{A}_s$ with greatest posterior probability $\mu_a(\boldsymbol{\beta}')$. This is then repeated until all $n$ runs are allocated.

There is again a key difference between the above allocation procedure and the standard Thompson sampling algorithm. The latter assumes rewards from different arms are independent, which may lead to slow exploitation of effective arms, since the procedure cannot borrow data from other arms. The proposed method, in contrast, leverages the fitted correlations from the Bayesian interaction model to guide allocation. By borrowing information from sampled arms, this then allows for quicker exploitation of effective arms given limited experimental data.




\subsection{Algorithm statement}

We now summarize the above parts into Algorithm \ref{alg:tsec}, which provides detailed steps for the TSEC procedure. TSEC begins with an initial selection of $K$ active arms $\mathcal{A}_s$, which can be chosen either uniformly-at-random or via an orthogonal array \citep{hedayat2012orthogonal}. Initial experiments are then performed by equally allocating $n/K$ runs for each arm in $\mathcal{A}_s$. Next, samples are drawn from the posterior distribution $[\boldsymbol{\beta}|\text{data}]$ using MCMC, then thinned to a set of $n$ samples $\boldsymbol{\beta}_1', \cdots, \boldsymbol{\beta}_n'$. For each thinned sample $j$, an experiment is then performed on the arm $a^*$ which maximizes the posterior probability $\mu_a(\boldsymbol{\beta}_j')$, $a \in \mathcal{A}_s$. Finally, the active arm set $\mathcal{A}_s$ is updated via \eqref{eq:ucb}. The last three steps are repeated sequentially until the experiment ends.


\begin{algorithm}[!t]
\caption{Thompson Sampling under Experimental Constraints}\label{alg:tsec}
\vspace{-0.2cm}
\bi
\setlength{\itemsep}{-5pt}
\item Initialize the $K$ arms in $\mathcal{A}_s$ for initial experimentation. 
\item Experiment on an equal allocation of $n/K$ runs on each arm in $\mathcal{A}_s$.
\ei
\vspace{-0.3cm}
\For{\textup{each switch period} $s = 1,2,...,S$}{%
\For{\textup{each time period} $t = 1,2,...,T$}{%
\bi
\setlength{\itemsep}{-5pt}
\vspace{-0.2cm}
\item Sample $\boldsymbol{\beta}_1, \cdots, \boldsymbol{\beta}_P$ from the posterior $[\boldsymbol{\beta}|\text{data}]$ using MCMC, where $P \gg n$.
\item Thin the sample chain into $n$ samples $\boldsymbol{\beta}_1', \cdots, \boldsymbol{\beta}_n'$.
\ei
 \vspace{-0.3cm}
\For{\textup{each experiment} $j = 1,2,...,n$}{%
\bi
\setlength{\itemsep}{-5pt}
\vspace{-0.2cm}
\item Experiment on the arm $a^* = \argmax_{a \in \mathcal{A}_s} \mu_a(\boldsymbol{\beta}_j')$ with largest sampled posterior probability, and observe binary reward $Y_{a^*} \in \{0,1\}$.
\vspace{-0.2cm}
\ei
}
}
\bi
\setlength{\itemsep}{-5pt}
\vspace{-0.2cm}
\item Update the active arm set $\mathcal{A}_s$ using \eqref{eq:ucb}.
\vspace{-0.2cm}
\ei
}

\end{algorithm}

\section{Simulated website optimization study}
We now investigate the effectiveness of TSEC in a simulated website optimization study. The simulation set-up and compared methods are chosen to mimic an actual website optimization study, which we unfortunately could not obtain since such data is highly proprietary.

\subsection{Set-up}
Here, we assume there are $M=10$ factors of interest, each with $L=2$ levels. These factors could represent font family, font size, image location, background color, etc. As discussed in Section \ref{sec:constr}, for website optimization, a company typically cannot afford to test out all website settings in an experimental period, due to high design and maintenance costs. To account for this arm budget constraint, suppose the experimenter can play only $K=16$ arms in each time period. In website optimization, the switching of arms typically happens at the end of the day when traffic is at its lowest. Within each experiment, we allow for $T=50$ time steps between each switch, each with $n=100$ experimental runs. A total of $S=5$ switches are allowed, yielding a total experimental time frame of $ST=250$ time periods.

We then simulate the reward generating mechanism from the assumed model \eqref{eq:model}. To set the true effect parameters $\boldsymbol{\beta}$, we adopt the framework in \cite{li2006regularities}, where effect parameters are populated from 113 published experiments across different engineering domains. The true main effect coefficients $(\beta_l^m)_{l=1}^{L_m}$ are simulated i.i.d. from the following spike-and-slab distribution \citep{ishwaran2005spike}:

\[
    (1-\gamma_{\rm ME})\mathcal{N}(0,1^2) + \gamma_{\rm ME}\mathcal{N}(0,10^2),
\]
where $\gamma_{\rm ME}$ is a Bernoulli random variable, with $p_{\rm ME} = \mathbb{P}(\gamma_{\rm ME} = 1) = 0.41$ being the probability that the main effect is significant. The true two-way interaction effects ${(\beta_{l,l'}^{m,m'})_{l=1}^{L_m}}_{l=1}^{L_{m'}}$ are simulated i.i.d. from the following distribution:
\[
    (1-\gamma_{\rm 2FI})\mathcal{N}(0,0.278^2) + \gamma_{\rm 2FI}\mathcal{N}(0,2.78^2),
\]
where $\gamma_{\rm 2FI}$ is a Bernoulli random variable, with $p_{\rm 2FI} = \mathbb{P}(\gamma_{\rm 2FI} = 1) = $
being the probability of an interaction effect being significant. Here, we wish to also incorporate the intuitive principle of effect heredity \citep{WuHamada2009}, which states that an interaction is active only when one or both of its parent effects are active. This can be incorporated into the sampling probability $p_{\rm 2FI}$ as:
\[
    \label{p2}
    p_{\rm 2FI}= 
\begin{cases}
    0.33,& \text{if both parent effects are significant,}\\
    0.045, & \text{if one of the parent effects is significant,}\\
    0.0048, & \text{if none of parent effects are significant.}
\end{cases}
\]
Finally, the true three-way interactions are simulated i.i.d. from the following distribution:
\[
    (1-\gamma_{\rm 3FI})\mathcal{N}(0,0.137^2) + \gamma_{\rm 3FI}\mathcal{N}(0,1.37^2),
\]
where $\gamma_{\rm 3FI}$ is a Bernoulli random variable, with $p_{\rm 3FI} = \mathbb{P}(\gamma_{\rm 3FI} = 1)$. To incorporate heredity, we sample the probability $p_{\rm 3FI}$ as:
\[
    \label{p3}
    p_{\rm 3FI}= 
\begin{cases}
    0.15,& \text{if all three parent effects are significant,}\\
    0.067, & \text{if two of the parent effects are significant,}\\
    0.035, & \text{if one of the parent effects is significant,} \\
    0.012,& \text{if no parent effects are significant.}\\
\end{cases}
\]

The proposed TSEC method is then compared with the following three industry benchmarks, which are commonly used for website optimization. Here, ``standard Thompson sampling'' refers to the existing Thompson sampling procedure in Section \ref{sec:ts}, with no modeled correlation between arms.
\begin{enumerate}
    
    \item \textbf{Benchmark 1}: The first baseline performs standard Thompson sampling within each experimental round. The active arm set $\mathcal{A}_s$ is chosen as a random $2^{10-6}$ factorial design \citep{mukerjee2007modern}, and is then fixed throughout the experimental period.
    
    \item \textbf{Benchmark 2}: The second baseline performs standard Thompson sampling within each experimental round. The active arm set $\mathcal{A}_s$ is initially chosen as a random $2^{10-6}$ factorial design.  At the end of each round, sub-optimal arms (arms with less than 5\% expected posterior probability of being the best arm) are discarded, and new arms are randomly added to the active arm set.
    
    \item \textbf{Benchmark 3}: The third baseline performs standard Thompson sampling within experimental round. The active arm set $\mathcal{A}_s$ is initially chosen as a random $2^{10-6}$ factorial design. At the end of each round, the top $K$ arms are selected based on their expected posterior probabilities from standard Thompson sampling.
    
\end{enumerate}

\begin{figure}[!t]
  \centering
  \includegraphics[scale = 0.7]{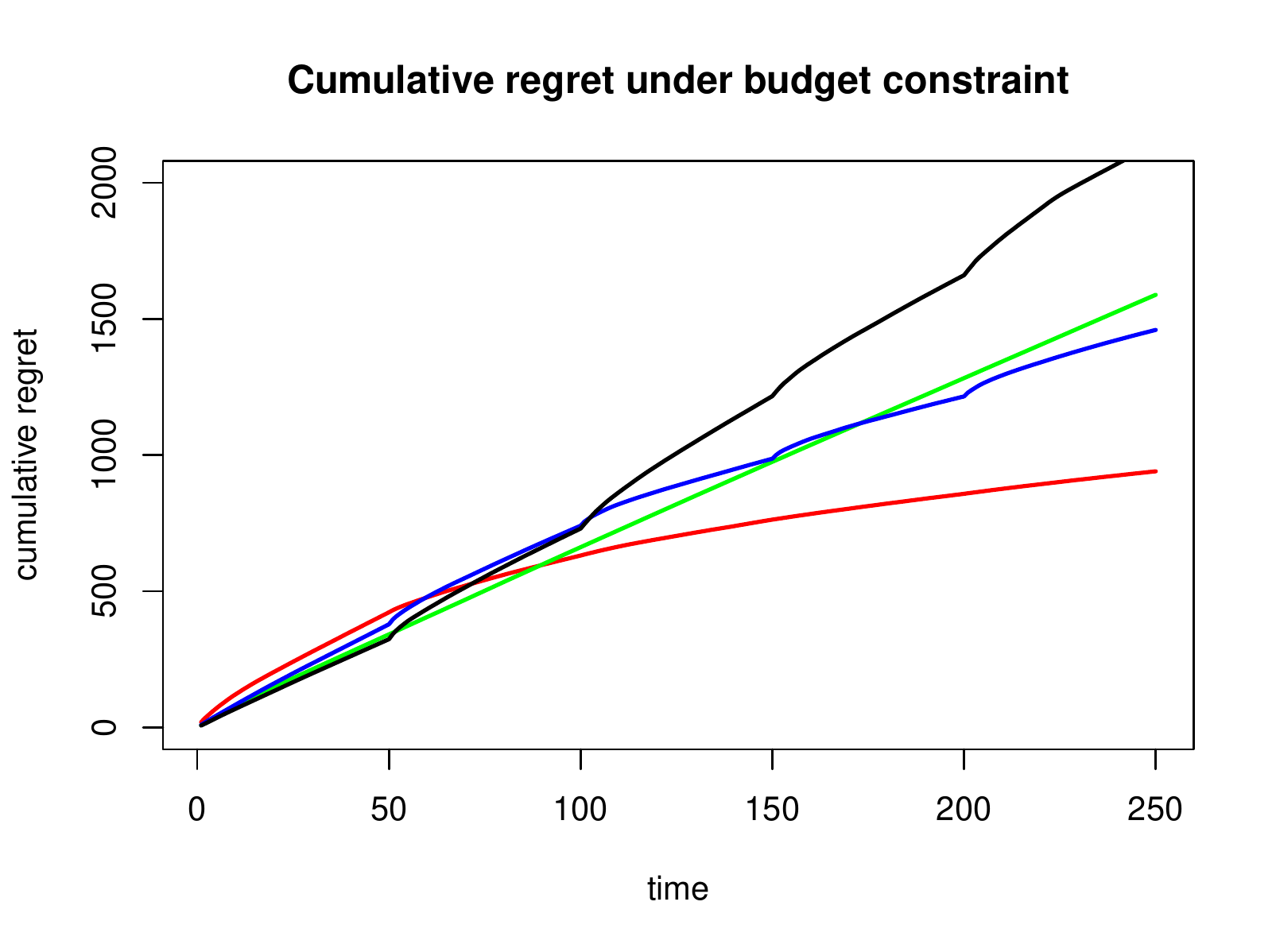}
  \includegraphics[scale = 1.0]{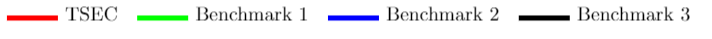}
  \caption{Cumulative regret plotted against experimental time period for the four considered website optimization methods.}
  \label{baseline-switch}
\end{figure}

\subsection{Analysis}

Figure \ref{baseline-switch} shows the cumulative regret (as defined in \eqref{eq:cumreg}) for the four considered methods over the experimental period. During the first switching period ($s=1$, times 1 - 50), we see that TSEC performs comparably to the three benchmarks. This is expected, since for all four methods, the initial active arms are selected via a factorial design. However, as the experiment progresses, the proposed method shows significantly lower cumulative regret over the three benchmarks. There are two reasons for this, both relating to the underlying fitted Bayesian interaction model. First, the correlations from this fitted model allow TSEC to add promising new arms and remove poorly performing arms during switches. Second, the posterior samples from the model then allow TSEC to allocate resources to promising arms, even if no data has been observed on such arms yet. At the end of the experiment, TSEC yields much lower cumulative regret compared to the three benchmarks, which shows the proposed algorithm outperforms these industry benchmarks by a noticeable margin.


Next, we investigate the performance of TSEC under the following modifications:
\begin{enumerate}
    \item We increase the total number of arms (i.e., website settings) from $N = 2^8$ to $N = 2^{15}$. This mimics a website optimization problem with many factors of interest and thereby many possible website designs.
    \item We change the arm budget constraint (i.e., the number of websites which can be tested in a given time period) from $K=16$ arms to $K=32$ and $K=64$ arms.
\end{enumerate}
\noindent These modifications will shed light when TSEC may perform well or poorly to the benchmark methods.


\begin{figure}[!t]
  \centering
  \includegraphics[scale = 0.7]{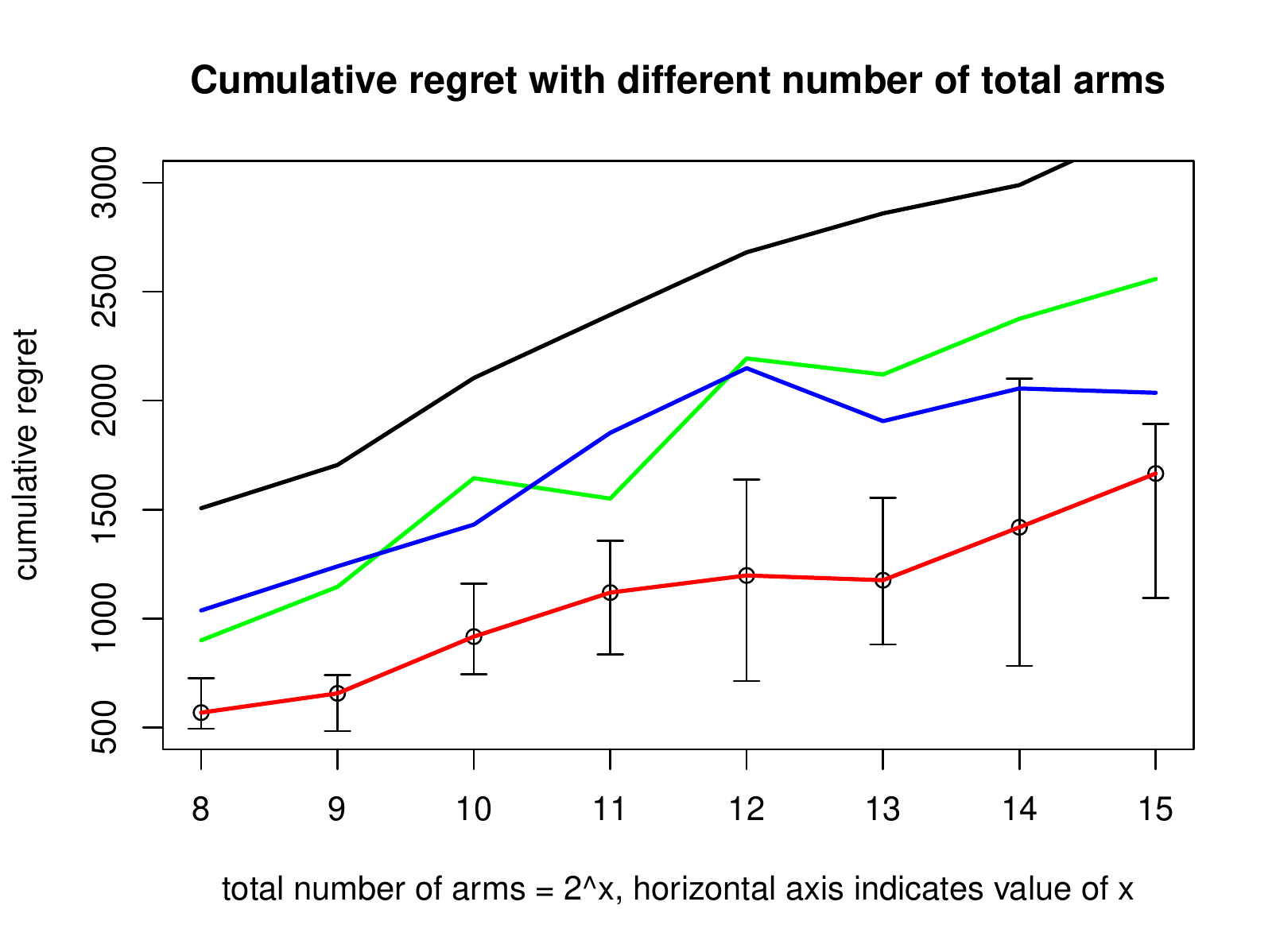}
  \includegraphics[scale = 1]{leg0}
  \caption{Cumulative regret (at the end of the experimental period) plotted against the total number of website settings $N$, for the four website optimization methods. Vertical bars show the 95\% confidence intervals for TSEC cumulative regret.}
  \label{total-switch}
\end{figure}

Figure \ref{total-switch} shows the cumulative regret for the four methods at the end of the experimental period, with the total number of website settings (arms) $N$ ranging from $2^8$ to $2^{15}$. There are a few interesting observations to note. First, as before, TSEC yields noticeably lower cumulative regret compared to the three existing benchmarks, which again shows the proposed method is quite effective at leveraging the fitted correlations from the Bayesian model for arm selection and resource allocation. Second, we see that the slope for TSEC in Figure \ref{total-switch} is smaller than the slopes for Benchmarks 1 and 3 (and comparable to the slope for Benchmark 2) as the number of arms increases. This suggests that the improvement of our method (compared to Benchmarks 1 and 3) grows as the design space of websites becomes larger. This is quite intuitive, since the fitted Bayesian interaction model to pool in data from different arms, which allows TSEC to infer reward rates over the high-dimensional arm space even with limited experimental data. Here, Benchmark 2 also yields a comparable slope to TSEC and a smaller slope than the other two benchmarks, but is noticeably worse than TSEC in terms of cumulative regret at all choices of $N$. This may be due to the random addition of new arms in Benchmark 2, which can be suboptimal when experimental resources are limited.

\begin{figure}[!t] 
  \Centering
     \includegraphics[width=.45\linewidth]{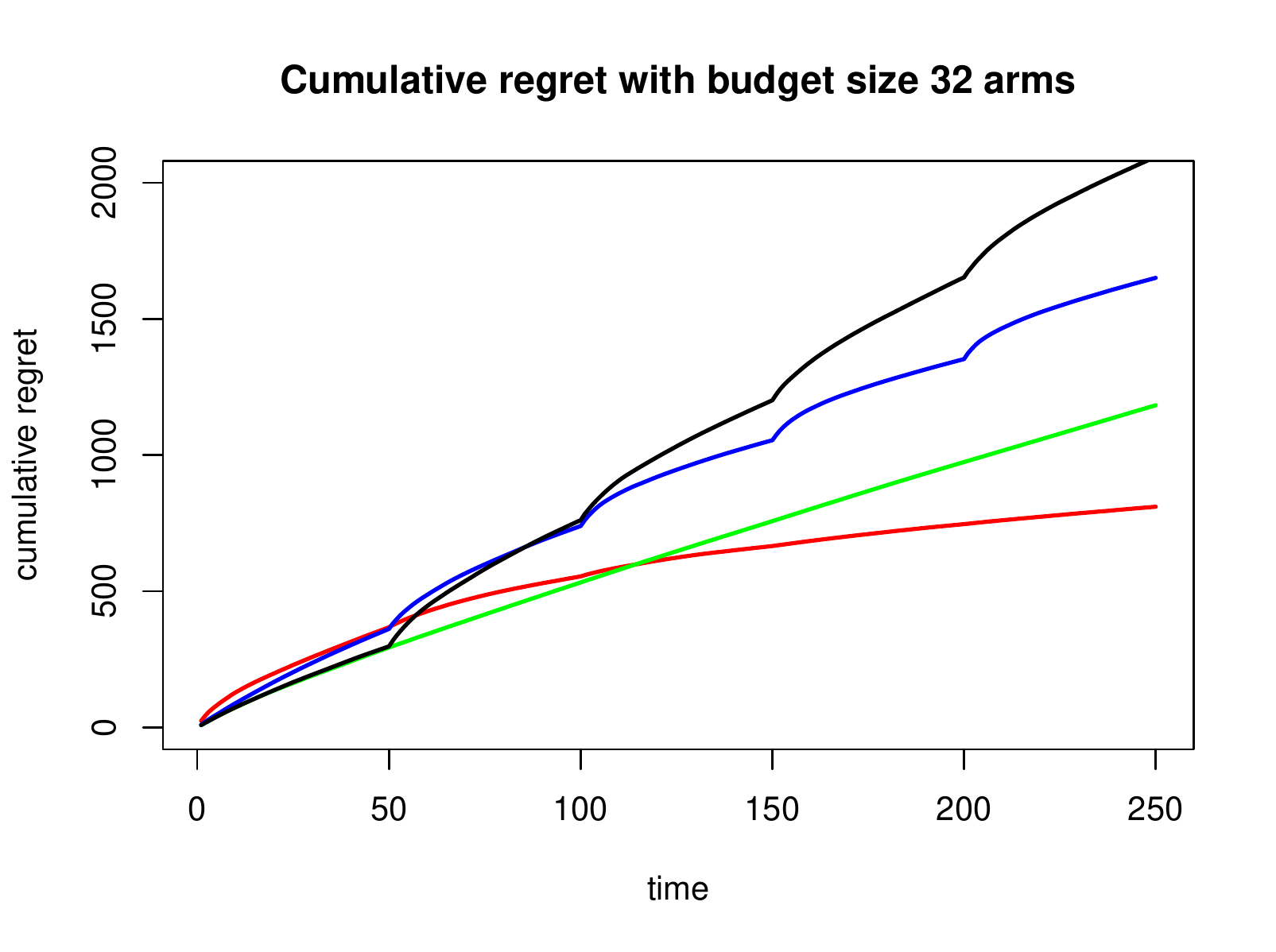}
     \includegraphics[width=.45\linewidth]{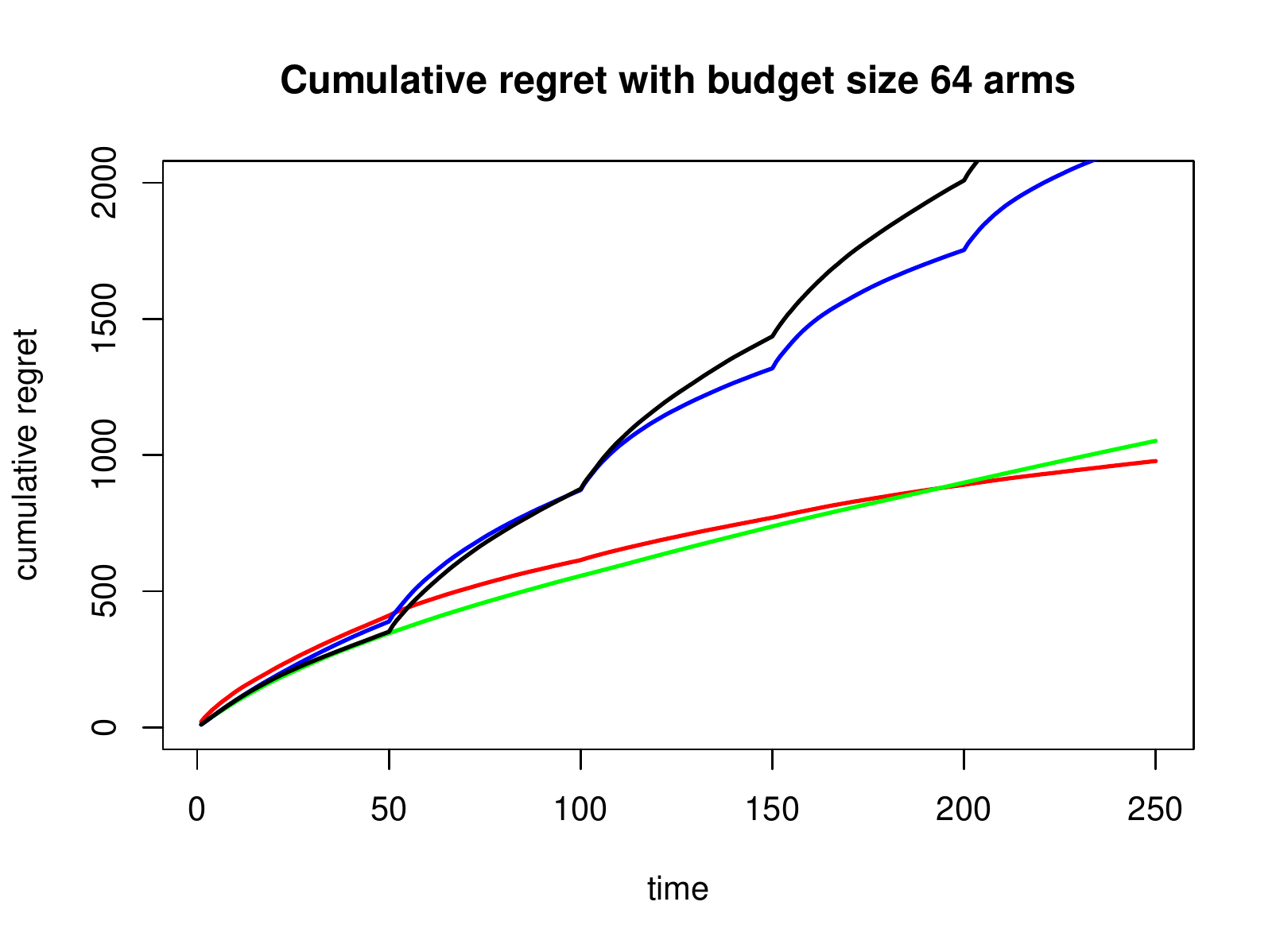}\\
     \includegraphics[scale=1.0]{leg0}
    \caption{Cumulative regret plotted against experimental time period under an arm budget constraint of $K=32$ (left) and $K=64$ (right), for the four website optimization methods.}
 \label{budget}
\end{figure}

Figure \ref{budget} shows the cumulative regret over time as we increase the arm budget constraint from $K=16$ to $K=32$ (left plot) and $K=64$ (right plot). We notice that, while TSEC maintains an advantage over existing benchmarks, its improvement over Benchmark 1 becomes smaller as the arm budget $K$ increases. This is again quite intuitive. Recall a key advantage of TSEC is that, given a limited arm budget and limited experiments, it can identify and exploit promising (but unobserved) arms via the fitted Bayesian interaction model. When this budget constraint is relaxed via a larger $K$, the value gained from this advantage decreases, which is as observed. Stated another way, TSEC yields greatest improvement over existing website optimization methods when there is a tighter experimental budget on websites tested within a time frame.


\section{Portfolio optimization application on ETFs}

Next, we investigate the performance of TSEC in a portfolio optimization application on exchange-traded funds (ETFs). We first provide a brief overview of the problem, then compare the performance of TSEC with standard investment strategies.


\subsection{Problem set-up}

Portfolio optimization \citep{black1992global} is a key topic of study in finance, and has profound impact on investment fund management and risk management \citep{brandt2010portfolio}. Loosely speaking, portfolio optimization is the process of optimizing one's holdings in a collection of assets to maximize capital returns. There are several well-established portfolio optimization strategies which have proven effective over the years. The Markowitz mean-variance framework \citep{markowitz}, which was awarded a Nobel Prize in Economics, is one of the most popular investment strategy used in practice. The core tenet of this strategy is to maximize expected returns of a portfolio at a fixed level of risk. Other popular investment strategies include the sold-all strategy, where an investor sells off the entire portfolio; the equally-weighted strategy, where an investor places equal weights on the considered assets; and the value-weighted strategy, where an investor places weights on each asset proportional to its returns in the previous investment period. Further details can be found in \cite{zhu2019adaptive}.

Any of the above four investment strategies can perform well (or perform poorly) under different market conditions. For example, in bullish market conditions where investor sentiment is generally positive, value-weighted strategies may be effective; otherwise, such strategies may perform poorly. These conditions may also vary greatly between industries: when certain industries are more bullish than others, an investor would want a value-weighted strategy which invests proportionally to the performance of that particular industry. It is therefore of great interest to have an \textit{online} portfolio optimization strategy, which can learn from market data effective combinations of investment strategies over different industries.


This problem can be nicely formulated into the considered multi-armed bandit framework. Suppose there are $M$ industries of interest. The space of industry-specific investment strategies can then be represented by the arm $\bm{x} = (x_1, \cdots, x_M)$, where $x_m \in \{1, \cdots, L=4\}$ encodes the investment strategy taken for industry $m$ (i.e., mean-variance, sold-all, equally-weighted, or value-weighted). In this sense, finding the best arm in this MAB is akin to finding the optimal industry-specific portfolio strategy. Furthermore, we encounter the same challenges which motivated the proposed TSEC method. First, the total number of industry-specific investment strategies is $N=4^M$; for a large number of industries $M$, this can be prohibitively large if experiments are performed on all arms. Second, transaction costs on each portfolio impose an arm budget constraint on this MAB problem. With these costs, it can be quite expensive to constantly rebalance different portfolios, which imposes a natural constraint on the number of arms (portfolios) played in a given period.


Since TSEC assumes that rewards are Bernoulli distributed, we would need to transform the observed portfolio returns to a binary reward variable, with 1 indicating success and 0 for failure. We follow the approach in \cite{zhu2019adaptive}, and use the Sharpe ratio \citep{sharpe1994sharpe} to encode the observed return. The Sharpe ratio is a measure of expected return per unit of risk, defined as:
\begin{equation*}
    \text{SR}(\text{portfolio}) = \frac{\mu(\text{portfolio})}{\sigma(\text{portfolio})},
\end{equation*}
where $\mu$ and $\sigma$ are the expected return and volatility of the portfolio. Both $\mu$ and $\sigma$ can be estimated from past market data (see \citealp{shen2016portfolio} for details). A reward of 1 is then assigned if the Sharpe ratio for the portfolio exceeds some pre-determined threshold $\tau$, and 0 otherwise. In our implementation, we used a Sharpe ratio threshold of $\tau=0.05$.

For data, we used the \texttt{quantmod} package \citep{ryan2008quantmod} in \textsf{R} to scrape market data from Yahoo Finance from January 1st, 2010 to August 31th, 2021. This encompasses a total investment period of 20 months. A total of $M=5$ industries are selected: healthcare, technology, finance, real estate, and gold, which yields a total of $N=4^5=1024$ different portfolio combinations. Within each industry, the top five exchange-traded funds (ETFs) are then selected based on their total assets under management. The adjusted closing price of the ETF is taken as the value of the stock for that day. We then take an arm budget constraint of $K=75$ portfolios to account for transaction costs, and allow for $S=20$ portfolio switches with $T=1$, which correspond to a rebalancing period of 30 days.

\subsection{Analysis}

\begin{figure}[!t] 
  \Centering
     \includegraphics[width=.7\linewidth]{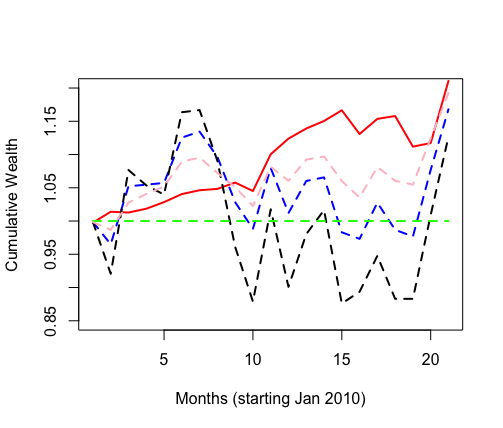}
     \includegraphics[scale=1.0]{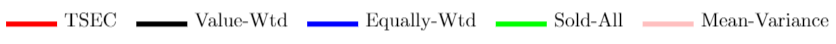}
    \caption{Cumulative wealth plotted against investment time (in months from January 2010) for TSEC and the four existing portfolio methods.}
 \label{fig:portfolio}
\end{figure}

Figure \ref{fig:portfolio} plots the cumulative wealth of the portfolio as a function of investment time in months, starting from January 2010, for TSEC and the existing four portfolio methods. We see that, initially, while the TSEC portfolio provides steady gains in cumulative wealth, it performs slightly worse than the mean-variance, equally-weighted, and value-weighted portfolios. This may be due to the initial \textit{exploration} of the TSEC method, where it investigates different industry-specific portfolio combinations (including suboptimal ones) to train the underlying Bayesian interaction model. After this learning period, however, TSEC yields higher cumulative wealth than the existing four methods. This may be due to the subsequent \textit{exploitation} of TSEC, where the proposed method makes use of the fitted Bayesian model to identify and experiment on promising industry-specific portfolio combinations. At the end of the 20-month investment period, the TSEC portfolio yields a return of over 20\%, which is noticeably higher than the existing four portfolio methods.

This figure also suggests an appealing quality of the TSEC portfolio, namely, its steady gains in cumulative wealth over the investment period. Note that the equally-weighted and value-weighted strategies, while yielding profit at the end of the investment period, can fluctuate greatly in the short-term, even yielding noticeable losses for the portfolios within the investment period. This would be undesirable for short-term investors, who have shorter investment horizons and may have to exit out of the investment while realizing capital losses. The mean-variance strategy is comparatively more stable, but yields lower cumulative wealth to TSEC. Compared to existing methods, TSEC shows a more steady increase in wealth, yielding consistent profits throughout the investment period. One reason may be that the proposed method provides better diversification of risk, since it explores a broader range of investment strategy combinations between industries. By modeling for the underlying correlation structure between arms, TSEC is able to explore this high-dimensional investment space, and identify effective portfolios under experimental constraints.

\section{Conclusion}

Online experimentation is a widely-adopted framework in a wide array of industrial applications, ranging from website design to portfolio optimization. In such applications, however, a practitioner is often limited by the number of choices (or arms) which can be experimented on, due to experimental costs. We propose a new method, called Thompson Sampling under Experimental Constraints (TSEC), which addresses this arm budget constraint in a multi-armed bandit setting. TSEC makes use of a Bayesian interaction model with effect hierarchy priors, to model correlations between rewards on different arms. We show how this fitted model can be integrated within a popular algorithm called Thompson sampling, to jointly identify good arms for experimentation and to allocate resources over these arms. We then investigate the performance of TSEC in two problems: a simulated website optimization application, and a portfolio optimization application on industry-based exchange-traded funds. Both applications demonstrate the effectiveness of TSEC over industry standards given experimental constraints.


Motivated by these encouraging results, there are several interesting avenues for future work. One direction is a ``hybrid'' TSEC framework which adaptively integrates rank- and model-based optimization. As explored in \cite{Wea1990} and more recently in \cite{2017ATM}, a rank-based optimization method may outperform a model-based method when significant interactions are present and an experimenter has limited data. This framework can allow for a more robust MAB method which can adaptively exploit within-arm dependencies from data. Another direction is the extension of TSEC for more general class of reward distributions (beyond the Bernoulli bandits studied in this work), which will allow for greater applicability of our method.

\medskip

\bibliography{mybib}
 
\end{document}